\documentclass{appolb}
\usepackage{graphicx}
\usepackage{amsmath,amssymb,bbm}
\usepackage{color}

\begin{document}
\title{\vspace*{-17mm}{\small \hfill UWThPh-2013-29}\vspace*{12mm}\\
Current status of constraints on the elements of the neutrino mass matrix%
\thanks{Talk presented at the XXXVII International Conference of Theoretical Physics\\``Matter to the deepest,'' Ustro\'n, 1--6 September 2013.}%
}
\author{Patrick Otto Ludl
\address{Faculty of Physics, University of Vienna,\\
Boltzmanngasse 5, A--1090 Vienna, Austria.}
}
\maketitle
\begin{abstract}
We analyse the mass matrix of the three light neutrinos
in the basis where the charged-lepton mass matrix is diagonal and
discuss constraints on its elements for the Majorana and the
Dirac case.
\end{abstract}
  
\section{Introduction}
The recent enormous improvement of our knowledge
of the neutrino oscillation parameters suggests a
detailed investigation of the current 
constraints on the neutrino mass matrix. Most of these
constraints depend on the assumed nature of
neutrinos (Dirac or Majorana).

The structure of this paper is as follows. After a brief general discussion of
the light-neutrino mass matrix in section~\ref{section-neutrino_mass_matrix},
we will investigate the implications of the currently available data on the Majorana neutrino
mass matrix in section~\ref{section-Majorana}. In section~\ref{section-Dirac} we will discuss
constraints on the neutrino mass matrix in the Dirac case. Finally we will conclude
in section~\ref{section-Conclusions}.

\section{The light-neutrino mass matrix}\label{section-neutrino_mass_matrix}
In this paper we assume that there are exactly three light neutrino mass eigenstates
with masses smaller than $\mathcal{O}(1~\mbox{eV})$, \textit{i.e.}\ we assume that there are no light
sterile neutrinos. By the term ``neutrino mass matrix'' we thus always mean the $3\times 3$ mass matrix of the
three light neutrinos.

If neutrinos are \textit{Majorana particles}, we assume that there is a (possibly effective) mass term
 \begin{equation}
 \mathcal{L}=-\frac{1}{2}\overline{\nu_L^c} M_\nu \nu_L + \mbox{H.c.} = \frac{1}{2}\nu_L^T C^{-1}M_\nu \nu_L + \mbox{H.c.},
 \end{equation}
where $M_\nu$ is a complex symmetric $3\times 3$-matrix. Such a mass term directly arises from the type-II seesaw mechanism
and can be effectively generated
via the seesaw mechanisms of type I and III.

If neutrinos are Dirac particles, \textit{i.e.}\ if the total lepton number is conserved, we assume the
existence of three right-handed neutrino fields $\nu_R$ leading to the mass term
 \begin{equation}
 \mathcal{L}=-\overline{\nu_R} M_D \nu_L + \mbox{H.c.},
 \end{equation}
where $M_D$ is an arbitrary complex $3\times 3$-matrix.

Before we can discuss any constraints on the neutrino mass matrix, we have to specify a basis in
flavour space.\footnote{Since the gauge interactions are flavour-blind, we a priori have the freedom
of performing arbitrary rotations in flavour space.} In models involving flavour symmetries, the chosen
matrix representations of the flavour symmetry group specify the basis. Since we will at this
point not assume any flavour symmetries in the lepton sector, we are free to choose a basis.
For simplicity we will always choose a basis in which the charged-lepton mass matrix is given by
 \begin{equation}\label{Mldiag}
 M_\ell = \mathrm{diag}(m_e,\,m_\mu,\,m_\tau).
 \end{equation}

\section{Constraints on the Majorana neutrino mass matrix}\label{section-Majorana}
\subsection{Parameterization of the Majorana neutrino mass matrix}
In the basis specified by equation~(\ref{Mldiag}) the Majorana neutrino mass matrix has the form
 \begin{equation}
 M_\nu = U_\text{PMNS}^\ast \mathrm{diag}(m_1,\,m_2,\,m_3) U_\text{PMNS}^\dagger,
 \end{equation}
where $U_\text{PMNS}$ is the lepton mixing matrix and the $m_i$ ($i=1,2,3$)
are the masses of the three light neutrinos. As any unitary $3\times 3$-matrix, $U_\text{PMNS}$
can be parameterized by six phases and three mixing angles. We will use the parameterization
 \begin{equation}
 U_\text{PMNS} = D_1 V D_2
 \end{equation}
with
 \begin{equation}
 D_1 = \mathrm{diag}(e^{i\alpha},\,e^{i\beta},\,e^{i\gamma})\quad
 \text{and}\quad
 D_2 = \mathrm{diag}(e^{i\rho},\,e^{i\sigma},\,1).
 \end{equation}
The phases $\alpha,\,\beta$ and $\gamma$ are unphysical since they may be eliminated by a suitable
redefinition of the charged-lepton fields. On the contrary, $\rho$ and $\sigma$ are physical in the case
of Majorana neutrinos and are therefore referred to as the Majorana phases.
$V$ denotes the well-known unitary matrix
 \begin{equation}
 V=\begin{pmatrix}
                  1 & 0 & 0 \\
                  0 & c_{23} & s_{23} \\
                  0 & -s_{23} & c_{23}
   \end{pmatrix}
   \begin{pmatrix}
                  c_{13} & 0 & s_{13}e^{-i\delta}\\
                  0 & 1 & 0 \\
                  -s_{13}e^{i\delta} & 0 & c_{13}
   \end{pmatrix}
   \begin{pmatrix}
                  c_{12} & s_{12} & 0 \\
                  -s_{12} & c_{12} & 0 \\
                  0 & 0 & 1
   \end{pmatrix},
 \end{equation}
where $s_{ij}\equiv\mathrm{sin}\,\theta_{ij}$ and $c_{ij}\equiv \mathrm{cos}\,\theta_{ij}$
are the sines and cosines of the three mixing angles, respectively. The phase $\delta$ is responsible for a
possible CP violation in neutrino oscillations (also in the Dirac case) and is therefore frequently
referred to as the Dirac CP phase.
\subsection{Upper and lower bounds on the absolute values of the elements of the neutrino mass matrix}
The fact that the neutrino masses are the \textit{singular values} of $M_\nu$
allows to derive a generic upper bound on the absolute values $|(M_\nu)_{\alpha\beta}|$. From
linear algebra it is known that the absolute value of an element of a
matrix is smaller or equal its largest singular value. For the neutrino mass
matrix this implies~\cite{Correlations}
 \begin{equation}
 |(M_\nu)_{\alpha\beta}| \leq \max_k m_k.
 \end{equation}
Since this bound is valid for \textit{any} matrix, it holds also for
Dirac neutrinos.
The strongest bounds on the absolute neutrino mass scale come from cosmology,
where the sum of the masses of the light neutrinos is usually constrained to be at most
of the order $\mathcal{O}(1\,\text{eV})$---see \textit{e.g.}\ the list of upper bounds in~\cite{PDG}.
From this we deduce the approximate upper bound $m_k\lesssim 0.3\,\text{eV}$ leading to
 \begin{equation}
 |(M_\nu)_{\alpha\beta}| \lesssim 0.3\,\text{eV}.
 \end{equation}

In~\cite{Correlations} also an analytical lower bound on the $|(M_\nu)_{\alpha\beta}|$
is provided.
Defining $a_k\equiv m_k |V_{\alpha k}| |V_{\beta k}|$ one can show that
 \begin{equation}\label{lowerbound}
 |(M_\nu)_{\alpha\beta}| \geq 2 \max_k a_k - \sum_k a_k. 
 \end{equation}
Note that this lower bound is independent of the Majorana phases $\rho$ and $\sigma$.
Unlike the generic upper bound discussed before, the lower bound~(\ref{lowerbound}) is
valid only for Majorana neutrinos. Numerically evaluating this lower bound using the
results of the global fits of oscillation data of~\cite{Forero,Fogli} only for two matrix elements
leads to non-trivial lower bounds. The lower bounds in units of eV for these
matrix elements are given by~\cite{Correlations}:
\vspace*{-2mm}
\begin{center}
\begin{footnotesize}
\begin{tabular}{lllccc}
 & & & $1\sigma$ & $2\sigma$ & $3\sigma$\\
$|(M_\nu)_{ee}|$ (inv. spect.) & Forero \textit{et al.} & \cite{Forero} & $1.52\times 10^{-2}$ & $1.36\times 10^{-2}$ & $1.14\times 10^{-2}$\\
                                           &  Fogli \textit{et al.} & \cite{Fogli} & $1.62\times 10^{-2}$ & $1.44\times 10^{-2}$ & $1.24\times 10^{-2}$\\
$|(M_\nu)_{\tau\tau}|$ (norm. spect.) & Forero \textit{et al.} & \cite{Forero} & $0$ & $0$ & $0$\\
                                                  &  Fogli \textit{et al.} & \cite{Fogli} & $1.86\times 10^{-2}$ & $1.27\times 10^{-2}$ & $0$\\
\end{tabular}
\end{footnotesize}
\end{center}
\medskip
For both global fits the only element being bounded from below at the $3\sigma$-level is $|(M_\nu)_{ee}|$
in the case of an inverted neutrino mass spectrum, for which in
both cases one finds $|(M_\nu)_{ee}|\gtrsim 10^{-2}\,\text{eV}$.
Unfortunately, this bound is still far from the current
upper bound stemming from searches for neutrinoless double beta decay,
which is given by~\cite{Rodejohann-betabeta,EXO-200}
\begin{equation}
m_{\beta\beta}\lesssim 0.4\,\text{eV}.
\end{equation}
\subsection{Correlations of the elements of the neutrino mass matrix}
In the case of Majorana neutrinos the absolute values of the elements of $M_\nu$ depend on
nine real parameters, namely
\begin{equation}
m_0,\, \Delta m_{21}^2,\, \Delta m_{31}^2,\, \theta_{12},\, \theta_{23},\, \theta_{13},\, \delta,\, \rho,\, \sigma,
\end{equation}
where $m_0$ denotes the mass of the lightest neutrino. Using the experimental/observational
constraints on these parameters, one can create plots of the allowed ranges of the $|(M_\nu)_{\alpha\beta}|$
versus $m_0$, which was first done by Merle and Rodejohann in~\cite{Merle-Rodejohann}.
In~\cite{Correlations} the analysis of~\cite{Merle-Rodejohann} was
repeated using the results of the recent global fits of oscillation data of~\cite{Forero,Fogli}. It turned out
that, at the $3\sigma$-level, the plots of~\cite{Merle-Rodejohann} are still in good agreement with the ones
of~\cite{Correlations}.

\begin{figure}[htb]
\centerline{%
\includegraphics[width=0.5\textwidth,angle=-90]{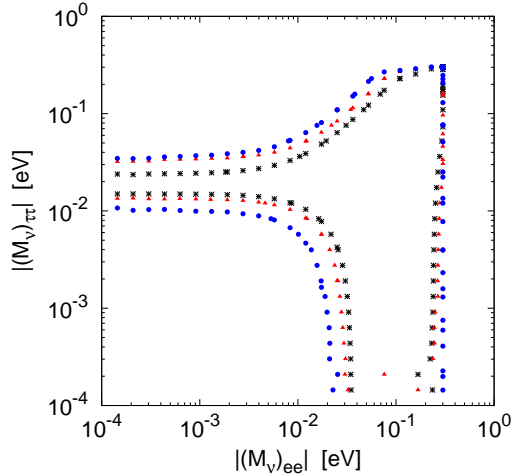}}
\caption{Correlation plot of $|(M_\nu)_{ee}|$ vs.\ $|(M_\nu)_{\tau\tau}|$ based on the global fit results of
Forero \textit{et al.}~\cite{Forero} assuming a normal neutrino mass spectrum and allowing $m_0$ to vary between
zero and $0.3\,\text{eV}$. The boundaries of the allowed areas are depicted by the following symbols:
best fit: {\large $\ast$},
$1\sigma:$ \textcolor{red}{$\blacktriangle$},
$3\sigma:$ \textcolor{blue}{$\bullet$}.}\label{Fig:M11-M33normal}
\end{figure}

In addition, in~\cite{Correlations} also
\textit{correlation plots} of the  $|(M_\nu)_{\alpha\beta}|$ were created. Since the Majorana neutrino
mass matrix has six independent entries, there are 15 correlations. Taking into account the two possible
neutrino mass spectra, there is a total of 30 plots. Among these 30 correlations one finds only five which
are manifest at the $3\sigma$-level, namely~\cite{Correlations}:
\begin{center}
\begin{tabular}{llll}
$|(M_\nu)_{ee}|$ & vs. & $|(M_\nu)_{\mu\mu}|$ & (normal spectrum)\\
$|(M_\nu)_{ee}|$ & vs. & $|(M_\nu)_{\mu\tau}|$ & (normal spectrum)\\
$|(M_\nu)_{ee}|$ & vs. & $|(M_\nu)_{\tau\tau}|$ & (normal spectrum)\\
$|(M_\nu)_{\mu\mu}|$ & vs. & $|(M_\nu)_{\mu\tau}|$ & (normal spectrum)\\
$|(M_\nu)_{\mu\tau}|$ & vs. & $|(M_\nu)_{\tau\tau}|$ & (normal spectrum).\\
\end{tabular}
\end{center}
All of these five correlations may be subsumed
as ``If one matrix element is small, the other one must be large.''
An example for such a correlation plot can be found in figure~\ref{Fig:M11-M33normal}.
In the case of an inverted neutrino mass spectrum, there are no correlations
manifest at the $3\sigma$-level.

It is important to note that while at the $3\sigma$-level the correlation plots
based on the global fits of~\cite{Forero} and~\cite{Fogli} agree, this is not true
at the $1\sigma$-level---for further details see~\cite{Correlations}.

\section{Constraints on the Dirac neutrino mass matrix}\label{section-Dirac}

\subsection{Parameterization of the Dirac neutrino mass matrix}

In analogy to the Majorana case, we will study the $3\times 3$ Dirac neutrino mass matrix
$M_D$ in the basis where the charged-lepton mass matrix is diagonal---see equation~(\ref{Mldiag}).
In this basis $M_D$ takes the form
\begin{equation}
M_D = V_R \mathrm{diag}(m_1,\, m_2,\, m_3) U_\text{PMNS}^\dagger,
\end{equation}
where $V_R$ is a unitary $3\times 3$-matrix. $V_R$ can be eliminated by
considering the matrix
\begin{equation}
H_D\equiv M_D^\dagger M_D = U_\text{PMNS} \mathrm{diag}(m_1^2,\, m_2^2,\, m_3^2) U_\text{PMNS}^\dagger.
\end{equation}
Since all observables accessible by current experimental scrutiny are contained in $H_D$,
all matrices $M_D$ leading to the same $H_D$ are \textit{indistinguishable} from the
experimental point of view. Therefore, the nine parameters of $V_D$ have to be treated
as \textit{free} parameters. Consequently, in stark contrast to the Majorana case, in the Dirac
case the neutrino mass matrix has at least nine free parameters (even if the mixing matrix and the neutrino
masses are known).

This freedom of choosing $V_R$ has important consequences for the analysis of $M_D$.
Obviously it is much harder to
put constraints on the elements of $M_D$ than in the Majorana
case. The freedom of choosing $V_R$ even allows to set several elements of $M_D$ to zero
without changing the physical predictions. This directly follows from the fact that
every matrix can be decomposed into a product of a unitary matrix and an upper
triangular matrix.\footnote{This factorization is known as the so-called QR-decomposition.}
Thus there is a choice of $V_R$ such that
\begin{equation}\label{triangular}
M_D = \begin{pmatrix}
m_{11} & m_{12} & m_{13}\\
0      & m_{22} & m_{23}\\
0      & 0      & m_{33}
\end{pmatrix}.
\end{equation}
Similarly, by multiplication of $M_D$ by one of the six $3\times 3$ permutation
matrices generated by
\begin{equation}
\begin{pmatrix}
0 & 1 & 0\\
0 & 0 & 1\\
1 & 0 & 0
\end{pmatrix}\quad\text{and}\quad
\begin{pmatrix}
1 & 0 & 0\\
0 & 0 & 1\\
0 & 1 & 0
\end{pmatrix}
\end{equation}
from the left, one can arbitrarily permute the rows of $M_D$ without changing any physical
predictions~\cite{Hagedorn-Rodejohann}.

However, it is most important to note that the freedom of choosing $V_R$ holds only
as long as we do not impose a symmetry in the lepton sector.
Namely, a flavour symmetry which acts non-trivially in the neutrino sector imposes constraints
on the form of the neutrino mass matrix $M_D$, \textit{i.e.}\ not only on the neutrino masses and
$U_\text{PMNS}$ but also on $V_R$. Consequently, a choice of $V_R$, \textit{e.g.}\ such that $M_D$ is upper
triangular, will in general be incompatible with the flavour symmetry.
Nevertheless, if we want to set bounds on the elements of $M_D$ (without introducing flavour symmetries), we indeed have the freedom
of arbitrarily choosing the matrix $V_R$.
Therefore, examining bounds and correlations of the elements of $M_D$ is much less elucidating
than in the Majorana case. That said, studies of $M_D$ such as the question
for the allowed cases of texture zeros in $M_D$ are still of great interest.

\subsection{Texture zeros in the Dirac neutrino mass matrix}

In the following we will shortly comment on the allowed cases of
texture zeros in $M_D$
under the assumption that $M_\ell$ is diagonal.
A detailed analysis has been done by Hagedorn and Rodejohann in~\cite{Hagedorn-Rodejohann},
which provides a classification of all possible texture zeros in this framework.
We repeated the analysis of~\cite{Hagedorn-Rodejohann} of the allowed cases of five, four and three
texture zeros\footnote{In~\cite{Hagedorn-Rodejohann} also the cases of one and two texture zeros
in $M_D$ are investigated, the result being that all these cases are allowed and do not show
any relations among the observables.} in $M_D$ based on the global fit results of~\cite{Fogli}.
Our numerical results are in perfect agreement with the analysis of~\cite{Hagedorn-Rodejohann}.
However, there are some previously allowed cases of texture zeros which can be excluded due
to the new data, namely
precisely those which lead to a vanishing or too
small value ($\lesssim 10^{-3}$) of $\mathrm{sin}^2\theta_{13}$,
\textit{i.e.}\ $A$, $B$, $\tilde{B}$, $C$, $D_1$--$D_3$, $\tilde{D}_1$--$\tilde{D}_3$, $E$
(inverted spectrum) and $\tilde{E}$ (inverted spectrum) in the notation of~\cite{Hagedorn-Rodejohann}.
Consequently, all cases of five texture zeros in $M_D$ are now excluded, and among
the cases of four texture zeros only $E$ (normal spectrum), $\tilde{E}$ (normal spectrum) as well as $F_1$--$F_3$
remain valid.

\section{Conclusions}\label{section-Conclusions}

In the case of Majorana neutrinos, the absolute values of the elements of the
light-neutrino mass matrix $M_\nu$ can be described by nine parameters, of which seven
are constrained by experiments/observations.
The by now very precise knowledge of the oscillation parameters therefore allows
detailed studies of the elements of $M_\nu$, including their allowed ranges and their
correlations.

The situation is quite different in the case of Dirac neutrinos, where the
neutrino mass matrix $M_D$ is by far not uniquely determined, even if the neutrino masses
and the mixing matrix are known. Therefore putting bounds on the elements of $M_D$
is much harder than in the Majorana case. Nevertheless studies of $M_D$ are possible, for example
the analysis of texture zeros in $M_D$. We reinvestigated the allowed texture zeros of $M_D$
in the basis where the charged-lepton mass matrix is diagonal. Our results agree with the
original analysis~\cite{Hagedorn-Rodejohann}, the only difference being that by now we know
that $\mathrm{sin}^2\theta_{13}\gg 10^{-3}$, which excludes some previously viable types
of texture zeros.

\bigskip
\noindent
\textbf{Acknowledgments:} The author wants to thank the organizers for the
invitation to the delightful and interesting conference.
This work is supported by the Austrian Science Fund (FWF), Project No.~P~24161-N16.

\end{document}